# TOWARD A HUMAN-CENTERED UML FOR RISK ANALYSIS
*Application to a medical robot*

Jérémie Guiochet[1], Gilles Motet[2], Claude Baron[2] and Guy Boy[3]
[1]*Grimm-Isycom/Lesia, Université de Toulouse II, 5 al. A. Machado, 31100 Toulouse, France;*
[2]*LESIA, INSA DGEI, 135 av. de Rangueil, 31077 Toulouse, France;* [3]*EURISCO International, 4 Av. E. Belin, 31400 Toulouse, France*

**Abstract**: Safety is now a major concern in many complex systems such as medical robots. A way to control the complexity of such systems is to manage risk. The first and important step of this activity is risk analysis. During risk analysis, two main studies concerning human factors must be integrated: task analysis and human error analysis. This multidisciplinary analysis often leads to a work sharing between several stakeholders who use their own languages and techniques. This often produces consistency errors and understanding difficulties between them. Hence, this paper proposes to treat the risk analysis on the common expression language UML (Unified Modeling Language) and to handle human factors concepts for task analysis and human error analysis based on the features of this language. The approach is applied to the development of a medical robot for tele-echography.

## 1. MOTIVATIONS

Today systems being more complex, and more responsibilities being transfered to them [1], safety requirement is becoming critical. Safety, previously defined as an absolute property [2], is also now expressed in a relative and probabilistic way as the property of a system to be "free from unacceptable risk" [3]. Therefore it is necessary to reduce the risk to an acceptable level with a complete risk management process [4], including activities presented on the left part of the figure 1. This approach has been used into different domains. For example, some of its concepts can be found in the medical standards [5]. Inside the general risk management activity, our study focuses on the first step: the risk analysis. This step aims at identifying hazards and estimating their associated risk (probability and severity). During this phase, various techniques can be used to handle functional and technological issues such as Preliminary Hazard Analysis (PHA), Failure Modes, Effects and Criticality Analysis (FMECA) and Fault Tree Analysis (FTA) techniques [6,7] as presented on the right part of figure 1.

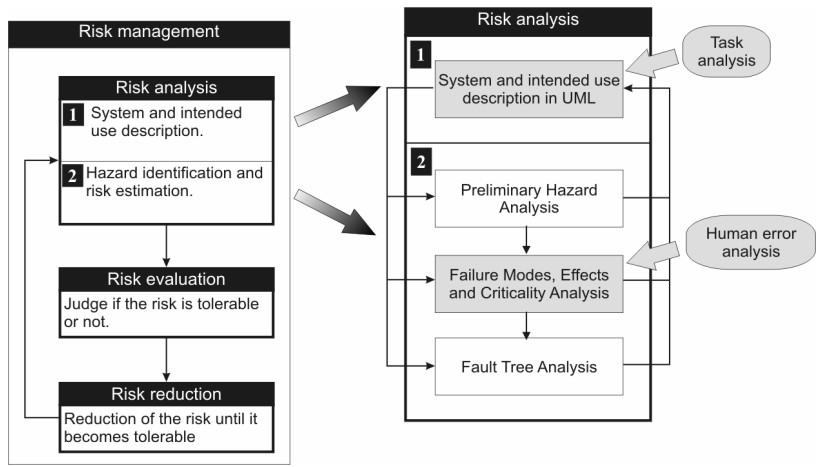

*Figure 1.* Human factors and UML based risk analysis in the risk management activity

The interaction between human and technological systems plays a major role in safety. Nevertheless, the integration of human factors in the risk management standards is still in work [8,9]. Based on this research, we focus on two main activities of human factors studies which are particularly important during risk analysis: "task analysis" for which the system and its intended use are described and "human error analysis" to identify new hazards and estimate their risks. These two phases are presented on the right part of figure 1. The second phase is implemented using FMECA [10].

Both activities are based on a system model. Ideally, the system definition is formally modeled. In practice, the use of formal methods in industrial development is still rare. A significant barrier is that many formal languages and analysis techniques are unfamiliar and difficult to apply for engineers. Moreover, several modeling tools have to be used to treat particular and partial aspects of the system. Designers must also communicate between specialists of different domains who usually have their own language. To handle this issue, we considered UML (Unified Modeling Language), which is now a standard in system and software engineering, even if this language presents several drawbacks (for instance, it has no formal semantics).

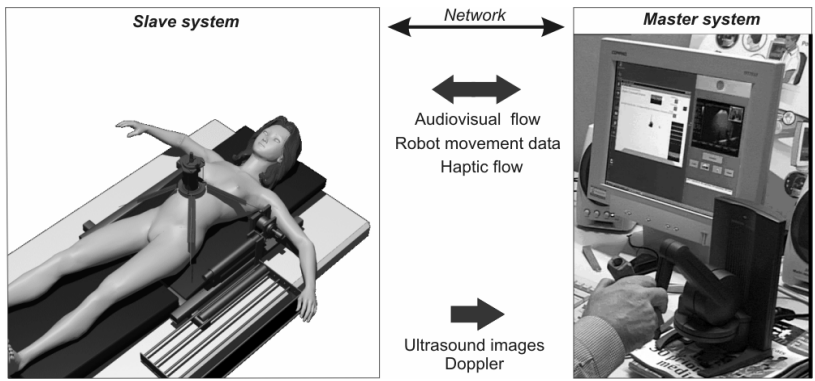

*Figure 2.* TER system overview

This paper presents how task analysis and human error analysis can be integrated in risk analysis and how UML can be useful to perform them. We will present some UML features to graphically specify tasks and to analyze them. Thus, section 2 exposes task analysis during system definition in UML and section 3 proposes an approach based on FMECA and UML message error models applied to human error analysis.

Each section is illustrated on a system for Robotic Tele-Echography (TER) [11]. TER is a tele-robotic system designed and developed by a French consortium composed of universities, hospitals and industrial companies. The slave robot is tele-operated by an expert clinician who remotely performs the ultrasound scan examination. A virtual probe is mounted on the master interface device. The real probe is placed on the slave robot end-effector. The slave robot is actuated with artificial muscles (pneumatic actuators). An overview of TER is provided on figure 2. We will focus on the computer control system of the slave site, whose safety is critical.

## 2. TASK ANALYSIS DURING SYSTEM DEFINITION WITH UML

Task analysis aims at identifying the details of specified tasks, including the knowledge, skills, attitudes, and personal characteristics required for successful task performance. During system analysis, this activity is linked to task allocation which aims at determining the distribution of work between human actors and machines. For instance, it is particularly important to define non ambiguous and consistent tasks for humans who are using the robot.

## 2.1   Related work

These activities are usually performed with different algorithms ([12,pp.231-236], [13] and [14]). Although there are a variety of techniques [15], the integration of task analysis in system modeling is still under development [16]. In this regard, many workshops aim at integrating human factors in system modeling [17], and more particularly in object oriented modeling [18,19]. Many studies compare *use cases* (based on Cockburn [20] definition which is closed to Jacobson's one [21]) and task analysis for interactive systems [22,23]. This was also applied for medical robots [24,25]. In those studies, use cases are usually derived from existing task analysis, and often led to the *business modeling* [26] like in [27]. Other authors study how to correlate task analysis and object oriented concepts, in order to model tasks themselves [28,29,30] for further human machine interface design.

Most of those theories are developed to design user interfaces. Our purpose is to provide a method to prevent hazards due to a bad task definition and allocation but also to provide models for human error analysis. This led us to analyze and model tasks with interaction diagrams, even if this "scenario-based" approach is sometimes opposed to "task-based" analysis as discussed in [31].

## 2.2   Business modeling

We first model the business without the technological system (ultrasound scan examination), with UML use cases and interaction diagrams (*collaboration* and *sequence diagrams*). During this step, "business modeling increases the understanding of the business and facilitates communication about the business" [26], particularly between engineers and doctors. For the considered example, the use case diagram in figure 3 models the common ultrasound scan examination without the robot system. Based on this diagram, the TER system is later integrated in the requirement modeling in the next diagrams.

Structuring the business with use cases helps the designers for the task allocation. For each use case, a textual description specifies more precisely the possible scenarios and their conditions of execution. Even if the use case *Perform Ultrasound Scan* seems to be the most important for the design, three other use cases have to be analyzed which can be later critical for the safety. Indeed, during the ultrasound scan examination the specialist can simultaneously manage the probe (change the settings), interact with the patient (communicate, prepare the surface to scan, etc.) and even diagnose. Hence, the future system should allow to perform all those use cases safely.

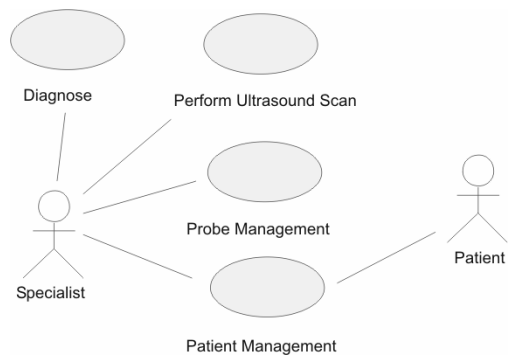

*Figure 3.* Use case diagram: global view of ultrasound scan examination

This notation completed with textual documents (or even task analysis studies) improves the communication between medical specialists and system designers. Based on this diagram, it is easier to allocate tasks between humans and systems. From the system analysis viewpoint, this documentation leads to design choices such as the system architecture presented on figure 2 (master and slave sites, bidirectional communication, etc.). Considering that the future system will integrate a robot, the other business is the use of a robot. This led to identify two generic use cases which are *Perform a task* and *Robot management*, but also two actors: the *Operator* and the *Robot* itself.

## 2.3 From business modeling to robotic system modeling

In this phase, we integrate the use cases of a robotic system into the use case diagram of the ultra sound scan examination (figure 3). This led to modify specifications of previous use cases. New actors are then identified. An *actor* characterizes an outside user or related set of users who interact with the system [4]. It is possible for an *actor* to be a human user (like the *Specialist* in the previous section) or an external system. This is really useful in socio-technical systems, and particularly in the TER project. We choose to represent two external systems as actors: the *Master Site* and the slave *Robot*. The *Master Site* replaces the actor *Specialist* (see figure 3) who is in charge of performing the examination. It is important to observe that the use case *Diagnose* has also disappeared, being transferred to the master site.

The use case diagram of figure 4 shows an allocation of work between actors. On this diagram, the boundaries of the computer control system are defined. It has been determined "which of the requirements are system requirements, which are requirements for the operational processes associated with the system and which requirements should be outside the scope of the system" [33]. This means that we have decided for each use case if it belongs or not to the system. For instance, the use case *Probe Management* has been removed from this use case diagram because it does not belong to or has any interaction with the computer control system.

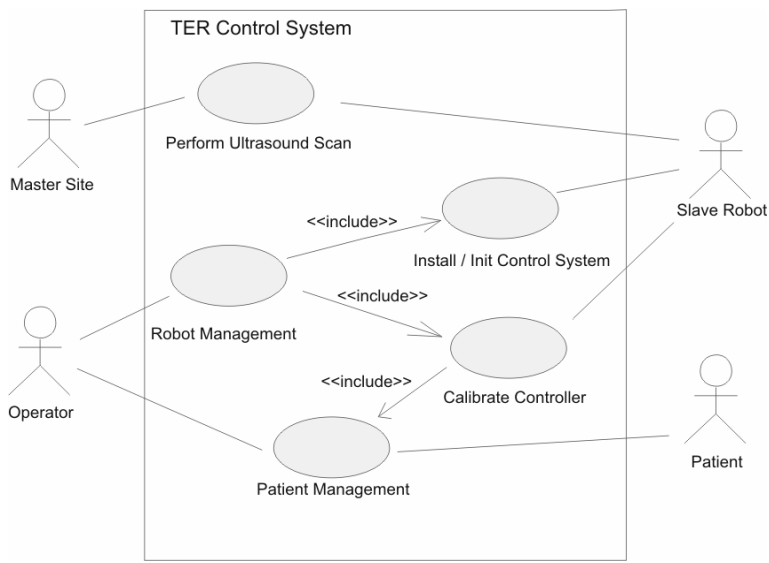

*Figure 4.* Use case diagram with Control System boundaries

## 2.4   Tasks description

We chose to specify tasks and subtasks with the UML concept of *message*. On sequence diagram figure 5, the main scenario of the use case *Install/Init Control System* is presented. This diagram can also be refined. For instance the *Operator* has to *Prepare Patient*, which can be detailed in: position the patient, put ultrasound scan gel on patient's body, give information to the patient, monitor the patient, etc. This notation of tasks is also useful to specify a sequence order, which can be essential for safety. By definition, sequence diagrams just specify possible scenarios (descriptive models). Nevertheless we use those diagrams as prescriptive models to establish a safe order of messages, because they are easily readable by non experts of UML modeling.

These models can directly be used for different safety-dependent tasks: writing of a user-guide (using the sequence diagrams), specifying and designing the Human-Machine Interface (HMI) and furnishing models for the specification of the system. It is important to note that in such robot systems, HMI includes the robot-human interface (control panels, teach pendant, etc.), but also the robot itself (in the TER project the slave robot is always in contact with the patient's body).

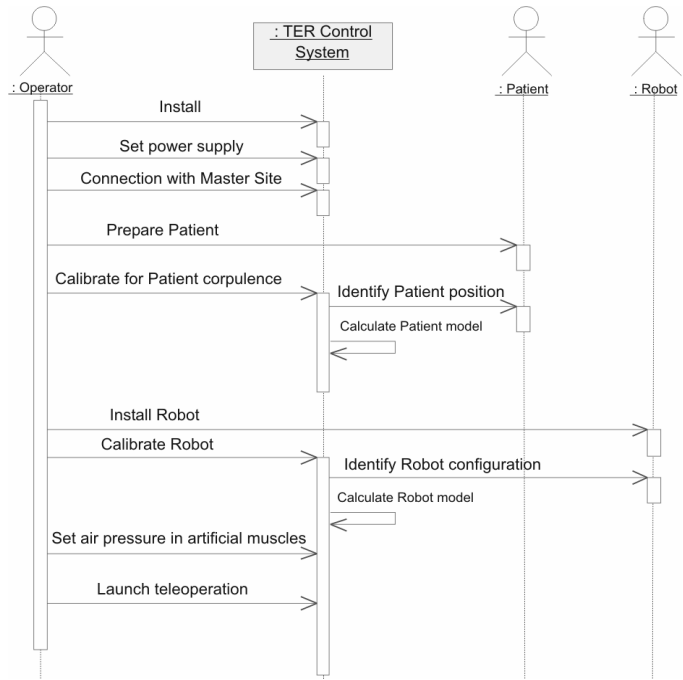

*Figure 5.* Sequence diagram of installation of the whole system

## 3. UML BASED FMECA FOR HUMAN ERROR ANALYSIS

As a potential source of harm (a hazard), human error has to be analyzed during the step of hazard identification and risk estimation. Although there is a variety of techniques (the most relevant and complete technique is certainly the Technique for Human Error Rate Prediction [34,35]) and tools [36], the complexity of human error classification and cognitive theory [37] usually leads engineers to the use of design checklists and guidelines [2] for the design of human computer interfaces. Nevertheless, as noted in [38], guidelines are not sufficient for innovative projects as medical robots.

## 3.1   Message failure mode analysis

The notion of failure mode is close to the notion of error; both concepts will be indifferently used in this section. In order to perform human error analysis, we have based our approach on several points. First, to be consistent with the previous section, we focus on a task-based analysis, quickly usable for non-specialist of human error. As proposed in [39], we do not have ethnographic studies and cognitive task analysis to perform the analysis. Thus, we only based our analysis on a set of models of scenarios, interface proposals, and human errors models.

Second, in order to reduce the number of analysis and modeling techniques, we propose to perform the human error identification and analysis with the well known analytical method FMECA. Among analytical methods allowing fault forecasting, FMECA [10] is certainly the most used during functional analysis. In an object oriented model, actors are represented as objects sending messages to the system. Hence, FMECA has to be conducted based on object concepts. This has been applied to object oriented elements such as components software [40], object methods [41], or use cases [42]. In those cases, the authors perform analysis focusing on functional aspects of object oriented models. On the contrary, the main idea of our approach is to propose an object oriented FMECA as we have previously done [43], and to apply it to objects such as actors.

The FMECA technique consists at first in identifying errors. These errors are often specific to the application. However, to realize a more systematic error identification step, one can sometimes use some generic error models. Those error models are related to generic elements of the system. In our approach we chose to focus on a central element of the UML dynamic diagram: the *message*. The concept of *Action* is also an important feature of UML to describe behaviors. But we did not handle this feature because its semantics changed a lot from version UML specification 1.4 [44] to 1.5 [45] and now to 2.0 [46].

## 3.2   Message error models

Most of language specifications contains *operational semantics* as well as *verification semantics*. The operational semantics is used to specify system functional aspects and to describe how the system will deliver the service. Most of UML diagrams belongs to operational semantics. The verification semantics defines properties associated with the correct use of features of the language. Some elements in the UML specification belong to the verification semantics. For instance, the use of constraints, graphically represented with curly brackets, allows to specify a restriction on a modeling element. There are also in UML the *Well-Formedness Rules,* which define a set of constraints expressed with the OCL language [44]. These constraint violations specify generic errors. However, most of verification properties are not explicitly provided in the UML standard. They are often implicitly integrated into the operational semantics, and thus they have to be deduced studying each feature of UML. As previously mentioned, our study focused on the *Message* feature.

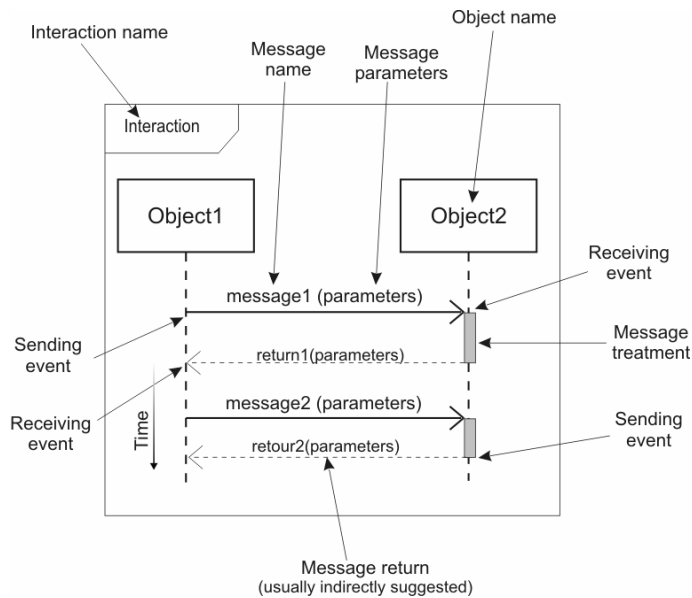

*Figure 6.* Elements of an interaction realized by the exchange of two messages

A message can be a signal creation, an operation call, a creation or destruction of an instance. The graphical representation by a sequence diagram is illustrated on figure 6. The different elements of a message are: the interaction it belongs, the next and previous messages in the interaction, the objects that send and receive the message, the sending and receiving events, the parameters (number, type and value), the implicit response (defined by its arguments, sending and receiving events), and the period of the message treatment.

Based on those elements we identify eleven error models (see [43] for details):

*E.1. Sending of a message not belonging to the planned interaction.*
*E.2. Execution of one or several messages in a wrong order.*
*E.3. Omission of a message among an interaction.*
*E.4. Lack of an instance to receive the message.*
*E.5. Sending or receiving of a message outside its specified time limits (too soon or too late).*

*E.6. The arguments type is different from the type of parameters expected by the receiver.*
*E.7. The number of message arguments is different from the number of parameters expected by the receiver.*
*E.8. The value of message arguments is different from the value of parameters expected by the receiver.*
*E.9. The values returned by a response to a message do not fit with the expected values (for example: constant, random, out of limits, etc.).*
*E.10. Treatment of a message out of the specified time limits.*
*E.11. Lack of link between sender and receiver objects.*

## 3.3 Proposition of a generic FMECA array for a system analysis

The error models been specified, their effects on system harm risk have to be studied. To handle this activity, we tuned the FMECA array [10] (originally devoted to functional analysis). This section proposes to introduce the following elements into the FMECA array for a message failure mode analysis (see figure 7): the interaction or the message name, the failure modes or the errors identified thanks to the previous error models, the causes of those failure modes, the effects at a local, higher or system level, the data to estimate the risk (*severity* is the harm seriousness, and failure mode occurrence is noted as *probability*), the on-line means to detect failure modes and their effects, the possible means for risk prevention and protection and other pieces of information.

Note that the goal in these arrays is not to proceed to a deep analysis of each of the mentioned points; in particular, the aim is not to consider the causes of the causes but to synthesize the main data in order to obtain a system analysis.

The *Potential solutions* of the array deal with the possible means to reduce the risk. It is important to notice that these means are not directly implemented but this highlights that a preliminary risk evaluation must be done. Risk is here calculated from a qualitative estimation of the probability of occurrence of a failure mode and of the severity of the induced harm. We chose to represent the prevention and protection means in order to reduce the probability or the severity of the considered harm.

This FMECA was essentially useful to focus on critical and weak design points from the safety point of view. Moreover, as FMECA directly depends on the model level of details, its use depends on the development process step. In our approach, we recommend to concentrate on the first steps, when safety requirements, architecture choices and major hazards are identified.

| Interaction/ Message | Failure mode (error) | Effects a. Same level b. Upper level c. System level | Risk Severity | Risk Probability | Risk | Possible detection means (online): a. Failure mode b. Effects | Potential solutions: a. Prevention b. Protection c. Other actions d. Remarks |
|---|---|---|---|---|---|---|---|
| Install/Init Control System:: **Set air pressure in artificial muscles** | Omission (E.3) | a. No power supply in artificial muscles b. No movements c. Patient waiting (stress) | 4 | P | I | b. Pressure sensor | a. Detailed user manual, formation, detailed actions on a screen b. Make a pumping test before launch teleoperation |
| | Wrong order (E.2) : before *Set power supply* | No initialization of Control System outputs. When power on: a. Spike of an output b. Uncontrolled movement c. Harmful movement for operator (patient not installed) | 2 | P | H | | a. Detailed user manual, formation, detailed actions on a screen b. Intelock system (to be defined) |
| | Pressure too high (E.8) | a. Reach the limit of intensity/pressure converters (to be determined) b. Partial or complete destruction c. Uncontrolled and harmful movements for patient | 1 | O | H | a. The operator check the pressure on a manometer | a. Indications on the manometer (close to the button) b. Pressure regulators before artificial muscles |

*Figure 7.* Example of a table of FMECA for the message
"Set air pressure in artificial muscles"

## 3.4 Application to the analysis of messages sent by actors

This section presents an example of use of error models previously identified (section 3.2) in order to demonstrate the tractability of the analysis proposed in section 3.3 for human error analysis. This approach has been successfully applied to medical robot system TER as presented in [6].

### 3.4.1 Types of errors

Merely all the error models previously identified can be applied for human error analysis. Common errors are the occurrence of an action of the actor not belonging to the planned interaction (error E.1), the execution of actions in a wrong order (E.2), and the omission of an action during an interaction (E.3). It is also possible to note human errors such as E.6, E.7 and E.8 consisting in furnishing bad data to the system. For instance, a user can type a letter whereas the system is waiting for a number (E.6), or he/she can tune a pressure valve too high for the system (E.8). The error E.4 is rather rare in human error analysis because it implies that the object for the interface is absent. The error E.5 depends on the time constraints a system can have, and is based on non functional requirements as for the error E.10. The error model E.9 which concerns response of a message (return values) is really useful for software or electronic components analysis. In case of the human component this is equal to E.6, E.7 and E.8, for message coming from the system.

### 3.4.2      Failure mode analysis

The modeling of all exchanged messages with UML sequence diagrams allows to perform an analysis very soon in the development. It can lead to formulate safety requirements from the start, without detailing design choices. Types of error are integrated in tables of an FMECA analysis in the column "Failure modes". For instance, we consider the message *Set air pressure in artificial muscles* from figure 5. As shown in figure 7, we identify three failure modes (the number has been reduced to present this example) from error models. In order to determine other columns data, we have to study all the UML models such state diagrams and class diagrams. Those diagrams are not presented here but can be found in [6,25]. For instance, those diagrams are used to determine effects of the failure modes on actors (column "Effects"). We proposed to use a scale for harm severity with five levels: negligible (5), minor (4), major (3), sever (2), catastrophic (1). Then, during a FMECA, it is easier to estimate the probability of the failure mode leading to the harm rather than the probability of the harm itself. Considering that a quantitative evaluation of the probability of occurrence of a human error is impossible to perform, we only do a qualitative estimation with different levels of probability of occurrence: frequent, probable, occasional, rare, and impossible. This point has to be developed, and relied to our type of errors. We have determined types of human errors that can appear in a human-machine interaction, but the causes are not integrated. In this table it is possible to highlight some important data missing for the analysis (like the maximum limit of air pressure in the converters).

## 4.      CONCLUSIONS AND PERSPECTIVES

The risk analysis approach proposed in this paper is motivated by the growing system complexity and safety requirements. At this stage, the human-centered approach of UML is twofold: a scenario-based task analysis and a message-based human error analysis.

We have shown that a scenario-based analysis performed throughout use case modeling helps designers in describing tasks. UML diagrams are initiated by UML specialists and further proposed to the other actors of the development process. This approach leads to a more consistent task allocation and to produce models that are useful in subsequent development steps.

Eleven error models have been presented. They are related to the concept of message in UML. In this paper, the object-oriented approach is linked to the FMECA functional risk analysis technique. Error models are integrated into FMECA. The resulting approach enables the various actors of the development process to use the same models.

This approach was applied successfully to the development of a first prototype of a medical robot for tele-echography. Others studies will be performed in different fields to complete and validate this work. The next technical step would be the development of tools to automatically integrate FMECA to UML design diagrams. We also need to go further in human error modeling to provide diagrams to understand how our types of error can be generated. Finally, a complete error model associated with the UML features is under development.